\documentclass[preprint,pre,tightenlines,aps]{revtex4-1}

\usepackage{amsmath}    
\usepackage{graphicx}   
\usepackage{verbatim}   
\usepackage{color}      
\usepackage{subfigure}  
\usepackage{hyperref}   
\usepackage{graphicx}
\usepackage{dcolumn}
\usepackage{bm}
\begin{document}

\title{Master regulators as order parameters of gene expression states}
\author{Andreas Kr\"amer}
\email{andreas.kramer@qiagen.com}
\affiliation{QIAGEN, Redwood City, CA 94063}

\date{\today}

\begin{abstract}
Cell type-specific gene expression patterns are represented as memory states of
a Hopfield neural network model. It is shown that order parameters of this model
can be interpreted as concentrations of master transcription regulators that
form concurrent positive feedback loops with a large number of downstream
regulated genes. The order parameter free energy then defines an epigenetic
landscape in which local minima correspond to stable cell states. The model is
applied to gene expression data in the context of hematopoiesis.
\end{abstract}

\maketitle

\section{Introduction}

The distinct cell types found in multi-cellular organisms exhibit characteristic
gene expression profiles that are generally viewed as being associated with
stable attracting states supported by the underlying gene regulatory network
\cite{Huang2005}. It has been proposed, that the Hopfield model
\cite{Hopfield1982} can be used to describe these cell type-specific expression
patterns in terms of memory states \cite{Lang2014, Fard2016, Guo2017}, however,
it is not a priori clear how such a model could be realized in a biological
context. It is shown here, that the Hopfield model emerges as an {\em effective}
model from a simple mechanism involving positive feedback loops with master
transcription regulators.

In biology, master regulators (MRs) are defined as transcription factors that
drive cell fate decisions \cite{Davis2017}. MRs are important in the context of
cancer \cite{Califano2016}, and essential for cellular reprogramming
applications \cite{Smith2016}. In general, they regulate expression of a large
number of downstream genes, and are observed to be mutually antagonistic in
different cell lineages \cite{Heinaniemi2013}, which suggests that their
abundance must be highly sensitive to cell type-specific gene expression
patterns. This motivates the idea that MRs are involved in positive feedback
loops with the expression-regulated genes which in turn leads to stable cell
states. In the model proposed here, the observed antagonism between MRs emerges
indirectly because these genes are regulated concurrently. Likewise,
interactions between genes that will appear in the Hopfield model, arise as 
{\em effective} interactions induced by the MRs.

A concept frequently employed in the context of cell development is that of an
energy-like epigenetic landscape which guides cell state changes. This concept
was first introduced by Waddington \cite{Waddington1939} as a qualitative
picture (``metaphor''), and is usually portrayed as a two-dimensional surface.
In the approach presented here, the epigenetic landscape is the free energy as a
function of the order parameters (OPs) of the Hopfield model, i.e. it is defined
in a space whose dimension is the number of cell types to distinguish, and OPs
measure concentrations of MRs. Local minima in this multi-dimensional landscape
then represent the different cell types. 

The paper is organized as follows: In Section II it is shown that the Hopfield
model is equivalent to a description in terms of feedback loops involving MRs
concurrently regulating expression of a large number of genes. Section III
applies this model to the hematopoietic cell lineages, and constructs an
approximate epigenetic landscape from published gene expression data.

\section{Feedback loops and the Hopfield model}

In the following it is assumed that gene expression is binary, i.e. genes are
either ``on'' or ``off'' corresponding to open or closed chromatin
configurations that enable or disable transcription controlled by transcription
factors. It is furthermore assumed that cell types are statistical ensembles of
individual cells with slightly varying gene expression states, so fluctuations
are driven by ``biological'' noise. The on/off-expression state of a regulated
gene $i$ is represented by a spin variable $s_i\in\{-1, 1\}$ with $i=1,..,N$,
governed by a distribution $P(\{s_i\})=\frac{1}{Z}e^{-H}$, where $H$ is a
Hopfield Hamiltonian 

\begin{equation}
H = -\frac{1}{2}\sum_{\substack{i,j \\ i\neq j}}s_iJ_{ij}s_j
\end{equation}

with Hebbian couplings 

\begin{equation}
J_{ij} = \frac{1}{N}\sum_k\beta_k\xi_i^k\xi_j^k.
\end{equation}

The parameters $\xi_i^k\in\{-1, 1\}$ with $k=1,..,M$, represent the stored
expression pattern for cell type $k$, and $\beta_k>0$ are pattern-specific
coefficients with values large enough so that the Hopfield model is in the
retrieval phase \cite{note_1}. An equivalent formulation \cite{Amit1985a} is
obtained by introducing Gaussian auxiliary variables $\phi_k$, and writing the
partition function $Z = Z(\{\beta_k\}, \{\xi_i^k\}) = \sum_{\{s_i\}}e^{-H}$ as

\begin{equation}
\label{Z}
Z = Z_0\sum_{\{s_i\}}\int\prod_kd\phi_k\exp\left[-\frac{N}{2}\sum_k
\frac{\phi_k^2}{\beta_k}+\sum_k\sum_i\phi_k\xi_i^ks_i\right]
\end{equation}

where the prefactor is $Z_0 = \left(\frac{N}{2\pi}\right)^{M/2}\prod_k
\beta_k^{-1/2}e^{-\beta_k/2}$. Integrating out the spin variables $s_i$ leads to 

\begin{equation}
Z = Z_0\int\prod_kd\phi_k\exp\left[-NV(\{\phi_k\})\right]
\end{equation}

where the effective potential $V(\{\phi_k\})$ is given by

\begin{equation}
\label{potential}
V(\{\phi_k\}) = \frac{1}{2}\sum_k\frac{\phi_k^2}{\beta_k} - \frac{1}{N}\sum_i
\log 2\cosh\left(\sum_k\phi_k\xi_i^k\right),
\end{equation}

and the auxiliary variables $\phi_k$ are identified as OPs of memory states $k$.
In the mean-field approximation, $\phi_k$ are given by the minima of the
potential $V$.

\begin{figure}[t!]
\centering
\includegraphics[width=15cm]{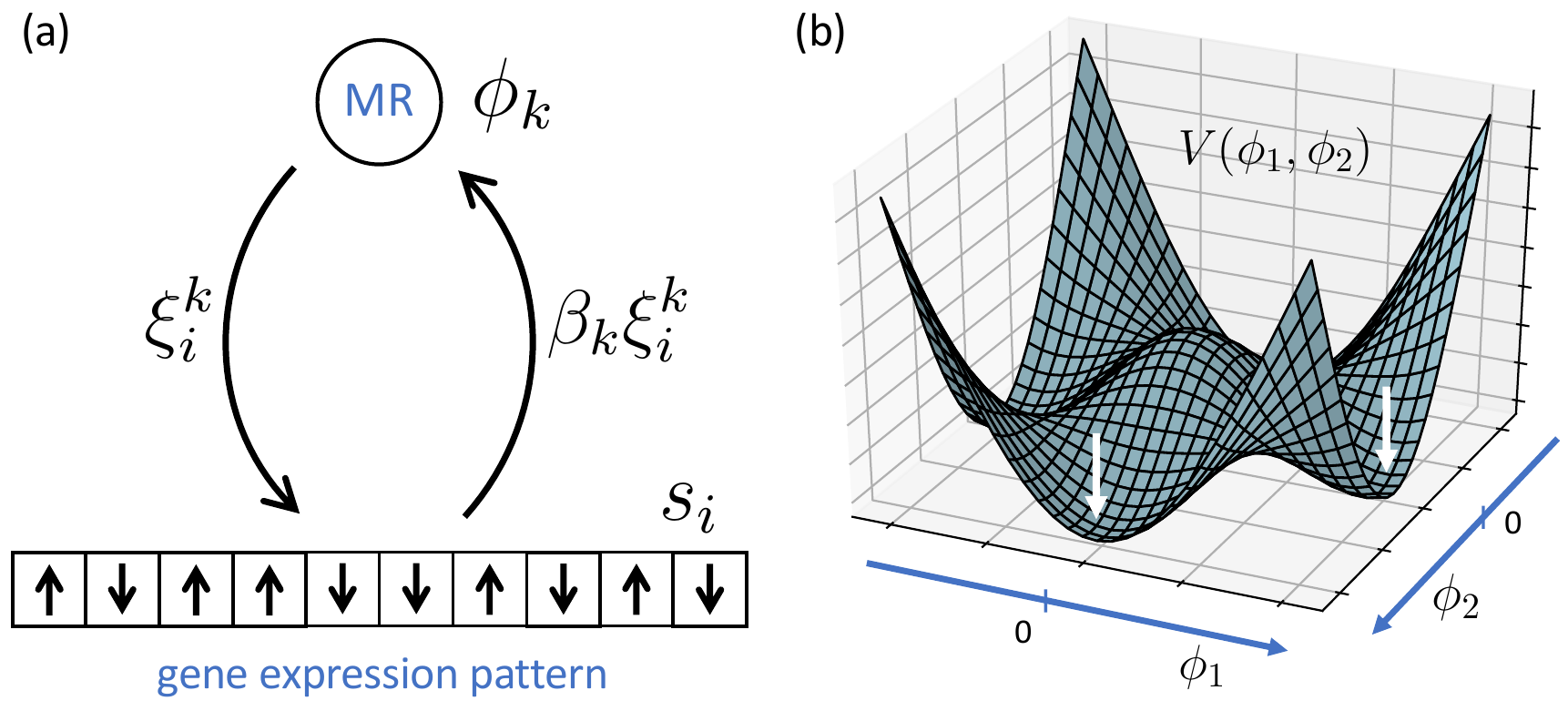}
\caption{\label{fig1} (a) Positive feedback loop involving master regulator (MR) 
and downstream regulated genes. MR concentrations are represented by the order 
parameter $\phi_k$, and gene expression states by the spin variables $s_i$. The 
forward and backward couplings of the feedback loop are $\xi_i^k$, and 
$\beta_k\xi_i^k$, where $\beta_k$ measures the strength of the feedback. 
(b) The ``epigenetic landscape'' $V(\phi_1, \phi_2)$ for the symmetric 
two-dimensional case ($\beta_1=\beta_2=2$). The biologically meaningful region 
is restricted to the positive sector $\phi_k\geq 0$ with arrows indicating 
the stable cell states.}
\end{figure}

In the following, we will interpret the OPs $\phi_k$ as concentrations of MRs
that form positive feedback loops with the expression-regulated genes, as is
schematically shown in Fig.~\ref{fig1}a. This is motivated by the form of the
forward and backward conditional probabilities derived from the joint
probability function underlying Eq.~(\ref{Z}),

\begin{equation}
\label{forward}
P\left(s_i\mid\{\phi_k\}\right) \sim \exp\left(s_i\sum_k\phi_k\xi_i^k\right),
\end{equation}

and 

\begin{equation}
\label{backward}
P\left(\phi_k\mid\{s_i\}\right) \sim \exp\left[-\frac{N}{2\beta_k}
\left(\phi_k-\frac{\beta_k}{N}\sum_is_i\xi_i^k\right)^2\right].
\end{equation}

Eq.~(\ref{forward}) shows $\phi_k$ as fields coupling to the spin $s_i$.
Assuming $\phi_k\geq 0$, this can be interpreted as MR $k$ acting as an
activator ($\xi_i^k>0$) or repressor ($\xi_i^k<0$) on gene $i$, i.e. pushing the
gene promotor state towards ``on'' or ``off'' in the cell ensemble.
Eq.~(\ref{backward}) in turn describes the feedback of the gene expression
pattern on the MR. For large $N$, the distribution
$P\left(\phi_k\mid\{s_i\}\right)$ is strongly peaked, so that the value of
$\phi_k$ is essentially a simple function of the overlap of the spin
configuration $s_i$ with the pattern $\xi_i^k$,
$\phi_k\approx\frac{\beta_k}{N}\sum_is_i\xi_i^k$. The feedback couplings
$\beta_k\xi_i^k$, where $\beta_k$ measures the strength of the feedback, are
proportional to the forward couplings, therefore a MR is sensitive to its own
regulation pattern, and orthogonal patterns have no effect. Note, that in both
cases the probabilities factorize, i.e. given $\{s_i\}$, the OPs $\phi_k$ are
independent random variables. Likewise, the spins $s_i$ are independent of each
other given $\{\phi_k\}$.

To be more specific, we can describe the feedback loop using first order 
kinetics in an idealized model,

\begin{equation}
\label{dynamics}
\frac{d\phi_k}{dt} = R_k - \phi_k
\end{equation}

where

\begin{equation}
R_k = \frac{\beta_k}{N}\sum_is_i\xi_i^k
\end{equation}

is the production rate of the MR controlled by the feedback mechanism, the
second term in Eq.~\ref{dynamics} describes its degradation, and the unit of
time has been set to 1. These dynamics assure that in the absence of other
forces the system is always driven to the equilibrium state $\phi_k^{eq} = R_k$.
Since $\phi_k$ is interpreted as a concentration, and $R_k$ is a production
rate, it shall always be assumed that $\phi_k\geq 0$ and $R_k\geq 0$.

For large $N$, ensemble fluctuations of $R_k$ can be neglected (they are of the
order $O(N^{-1/2})$), and $s_i$ can be replaced by its ensemble average
$\left<s_i\right>=\tanh\left(\sum_k\phi_k\xi_i^k\right)$ from
Eq.~(\ref{forward}). This leads to an expression of $R_k$ as a function of
$\{\phi_k\}$ alone,

\begin{equation}
\label{rate}
R_k = \frac{\beta_k}{N}\sum_i\xi_i^k \tanh\left(\sum_l\phi_l\xi_i^l\right),
\end{equation}

thus Eq.~(\ref{dynamics}) becomes

\begin{equation}
\frac{d\phi_k}{dt}=-\beta_k\frac{\partial V}{\partial\phi_k}.
\end{equation}

The potential $V(\{\phi_k\})$ in Eq.~(\ref{potential}) can therefore be
interpreted as an ``epigentic landscape'' driving the dynamics of MR
concentrations $\phi_k$. In the following - assuming that $N$ is large - the sum
$\frac{1}{N}\sum_i$ over expressions involving the patterns $\xi_i^k$ is
replaced by a ``quenched'' expectation value
$\langle\!\langle\cdot\rangle\!\rangle$ over the random variable $\xi^k$. It
shall be noted that it is not necessary to assume that the backward couplings
are strictly proportional to the forward couplings, since $\xi_i^k$ only appears
in averages over all spins. For instance, let $\tilde\xi_i^k=\eta_i^k\xi_i^k$,
where $\eta^k$ is a random variable that is sufficiently uncorrelated with
$\xi^k$ with $\langle\!\langle\eta^k\rangle\!\rangle>0$, then
$\langle\!\langle\tilde\xi^k\cdot\rangle\!\rangle\approx\langle\!\langle\eta^k
\rangle\!\rangle\langle\!\langle\xi^k\cdot\rangle\!\rangle$,
so that $\langle\!\langle\eta^k\rangle\!\rangle$ can be absorted in the feedback
strength $\beta_k$.

The second-order term in an expansion of the potential $V(\{\phi_k\})$ around 
$\phi_k=0$ is given by

\begin{equation}
\label{V2}
V^{(2)}(\{\phi_k\})=\frac{1}{2}\sum_{k,l}\left(\frac{1}{\beta_k}\delta_{kl}-
\langle\!\langle\xi^k\xi^l\rangle\!\rangle\right)\phi_k\phi_l.
\end{equation}

It is seen that the state $\{\phi_k\}=(0, 0, ..., 0)$ is stable if the feedback
strengths $\beta_k$ are small enough since the first term in Eq.~(\ref{V2}) is
dominating. When $\beta_k$ are increased, this state becomes unstable, and the
OPs $\phi_k$ are driven into other minima of $V$ away from zero. In the
symmetric case where the patterns $\xi_i^k$ are orthogonal with zero mean,
$\langle\!\langle\xi^k\xi^l\rangle\!\rangle=\delta_{kl}$ and
$\langle\!\langle\xi^k\rangle\!\rangle=0$, this transition happens at
$\beta_k=1$, and stable single-memory states are found to be of the form
$\{\phi_k\} \sim (0,...,0, 1,0,...,0)$. It is known that for larger values of
$\beta_k$ also mixtures of odd numbers of memories appear as ``spurious''
meta-stable states \cite{Amit1985a}. Because of the symmetry of the potential
$V$ w.r.t.~sign changes of $\phi_k$, in this case, the dynamics defined in
Eq.~(\ref{dynamics}) can be restricted to the biologically meaningful sector
$\phi_k\geq 0$ (note that sign changes of $\phi_k$ would correspond to a flip of
the corresponding spin pattern of the memory state). It then follows from
Eq.~(\ref{rate}) that the production rate $R_k$ is also always positive or zero:
$R_k=\beta_k\left<\!\left<\tanh\left(\sum_{l\neq
k}\phi_l\xi^k\xi^l+\phi_k\right)\right>\!\right>\geq\beta_k
\left<\!\left<\tanh\left(\sum_{l\neq
k}\phi_l\xi^k\xi^l\right)\right>\!\right>=0$, assuming mirror symmetry in the
probability distribution of $\xi^k$. For illustration, an example of the
potential $V$ is shown in Fig.~\ref{fig1}b for the two-dimensional case.

It shall be pointed out that several simplifying assumptions were made: The
combined effect of different MRs on the expression of individual genes is
assumed to be linear and additive, as is the combined effect of individual genes
on MRs for the feedback. The description of the system is highly idealized since
it neglects many details of gene transcription dynamics, protein translation,
regulation by post-translational modification, as well as the role of co-factors
and protein complexes. It is possible, that specific protein complexes may
actually be key to sensing particular patterns, since their formation is
sensitive to the concentration of individual protein components.

\begin{figure}[t!]
\centering
\includegraphics[width=15cm]{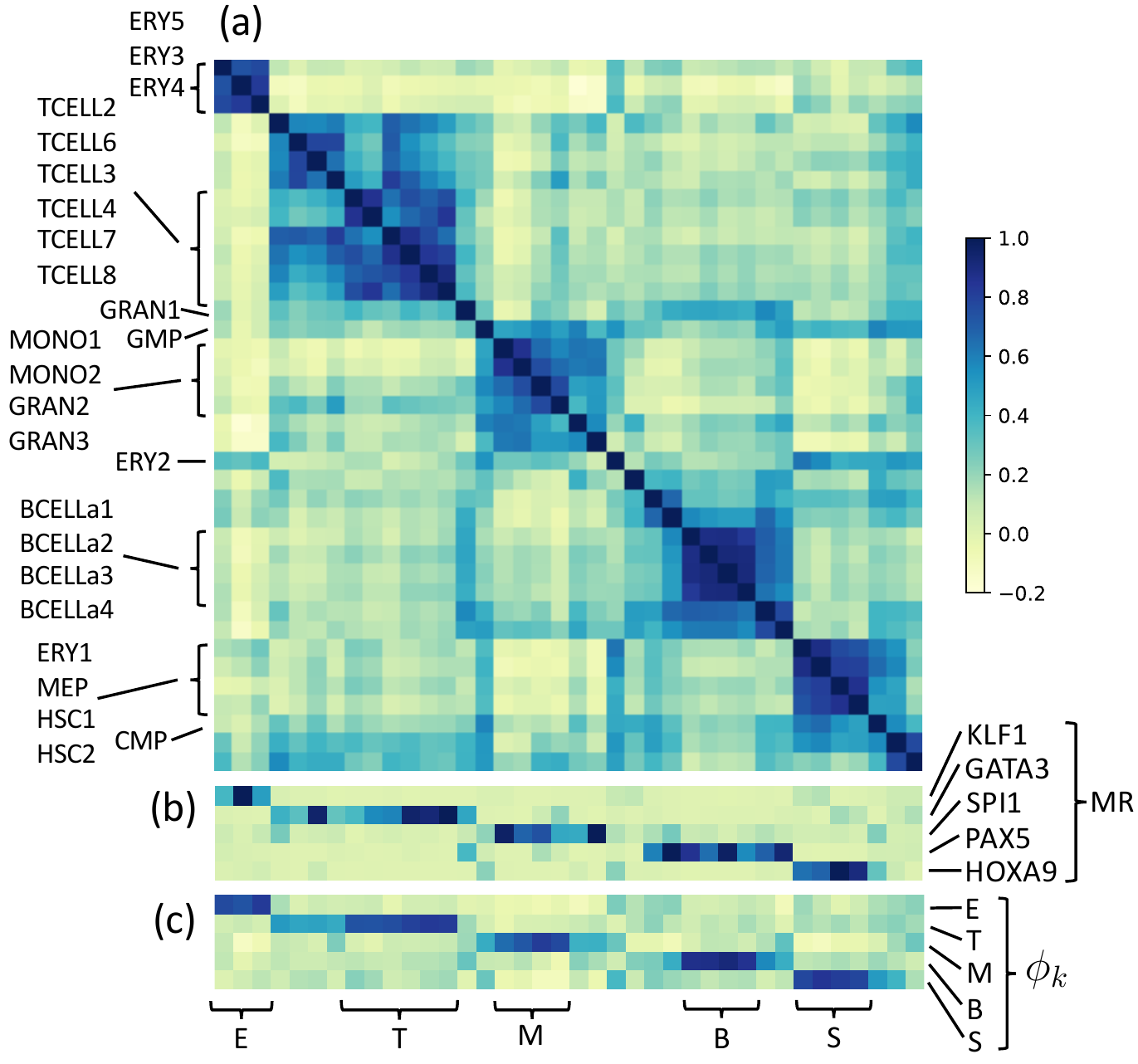}
\caption{\label{fig2} (a) Cell type correlation matrix based on expression 
states of $N=424$ selected genes as described in Section III.A. (b) Normalized 
expression for master regulators HOXA9, GATA3, PAX5, SPI1, and KLF1. (c) 
Corresponding computed order parameters for hematopoietic stem cell (S), 
T-lymphoid (T), B-lymphoid (B), myeloid (M), and erythroid (E) categories.}
\end{figure}

\section{Application to gene expression data}

In the following, the theoretical model described above is applied to gene
expression data in the context of blood cell development (hematopoiesis). Blood
cells form from hematopoietic stem cells (S) in the bone marrow into different
lineages of T-lymphocytes (T), B-lymphocytes (B), myeloid cells (M), and
erythroid cells (E). We assume that each of these cell categories, rather than
individual cell types, is represented by an OP $\phi_k$ governed by the
epigenetic potential $V(\{\phi_k\})$, where $k\in\{S, T, B, M, E\}$. The view is
that specific cell types are guided by the potential $V$, but are also subject
to perturbations that depend on the detailed biology, leading to OP values close
to the minima of $V$.  If for example the effect of these perturbutions is such
that a (small) fraction $\frac{n}{N}$ of the spins $s_i$ is fixed externally,
i.e. not subject to regulation by $\{\phi_k\}$, then the production rate $R_k$,
and hence the stationary state $\phi_k^{eq}$ is still given by the overlap of
the expression pattern, $\phi_k^{eq}=\frac{\beta_k}{N}\sum_is_i\xi_i^k$.
However, as is straighforward to show, the value of $\phi_k^{eq}$ will be
shifted due to a term added to the potential $V$, $\tilde
V=\frac{n}{N}\left<\!\left<\log
2\cosh\sum_l\phi_l\xi^l\right>\!\right>-\frac{n}{N}\sum_k\gamma_k\phi_k$, where
$|\gamma_k|\leq 1$.

\subsection{Analysis}

Gene expression data \cite{data} for 38 human hematopoietic cell populations
purified by flow sorting \cite{Novershtern2011} was mapped onto the interval
$[-1, 1]$ with -1 corresponding to ``not expressed'', 1 corresponding to
``expressed'', and intermediate values reflecting different levels of confidence
between these two limits. This mapping is motivated by the observation that
log$2$-scaled expression distributions for specific cell types generally exhibit
a bi-modal profile with two peaks that can be interpreted as ``on'' and ``off''
gene-promoter states \cite{Hebenstreit2014}. The map from log$2$-scaled gene
expression values $g_i$ to the interval $[-1, 1]$ was based on a ``soft'' sign
function around the sample median $\bar g$, $e_i=\tanh(g_i - \bar{g})$. In total
12,953 genes were included in the analysis. The different cell populations are
shown in Table~\ref{tab:table1}.

\begin{table}[b]
\caption{\label{tab:table1}%
The 38 hematopoietic cell populations from \cite{Novershtern2011} for which 
gene expression data was analysed.}
\begin{ruledtabular}
\begin{tabular}{ll}
\textrm{Symbol}&
\textrm{Description}\\
\colrule
HSC1,2 & hematopoietic stem cell\\
CMP & common myeloid progenitor\\
MEP & megakaryocyte/erythroid progenitor\\
ERY1-5 & erythroid cells\\
MEGA1,2 & megacaryocytes\\
GMP & granulocyte/monocyte progenitor\\
GRAN1-3 & granulocytes/neutrophils\\
MONO1,2 & monocytes\\
EOS & eosinophil\\
BASO & basophil\\
DENDa1,2 & plasmacytoid and myeloid dendritic cells\\
Pre-BCELL2,3 & B cell progenitors\\
BCELLa1-4 & B cells\\
NK1-3, NKT & NK cells\\
TCELL1,2-8 & T cells\\
\end{tabular}
\end{ruledtabular}
\end{table}

\begin{figure}[t!]
\centering
\includegraphics[width=12cm]{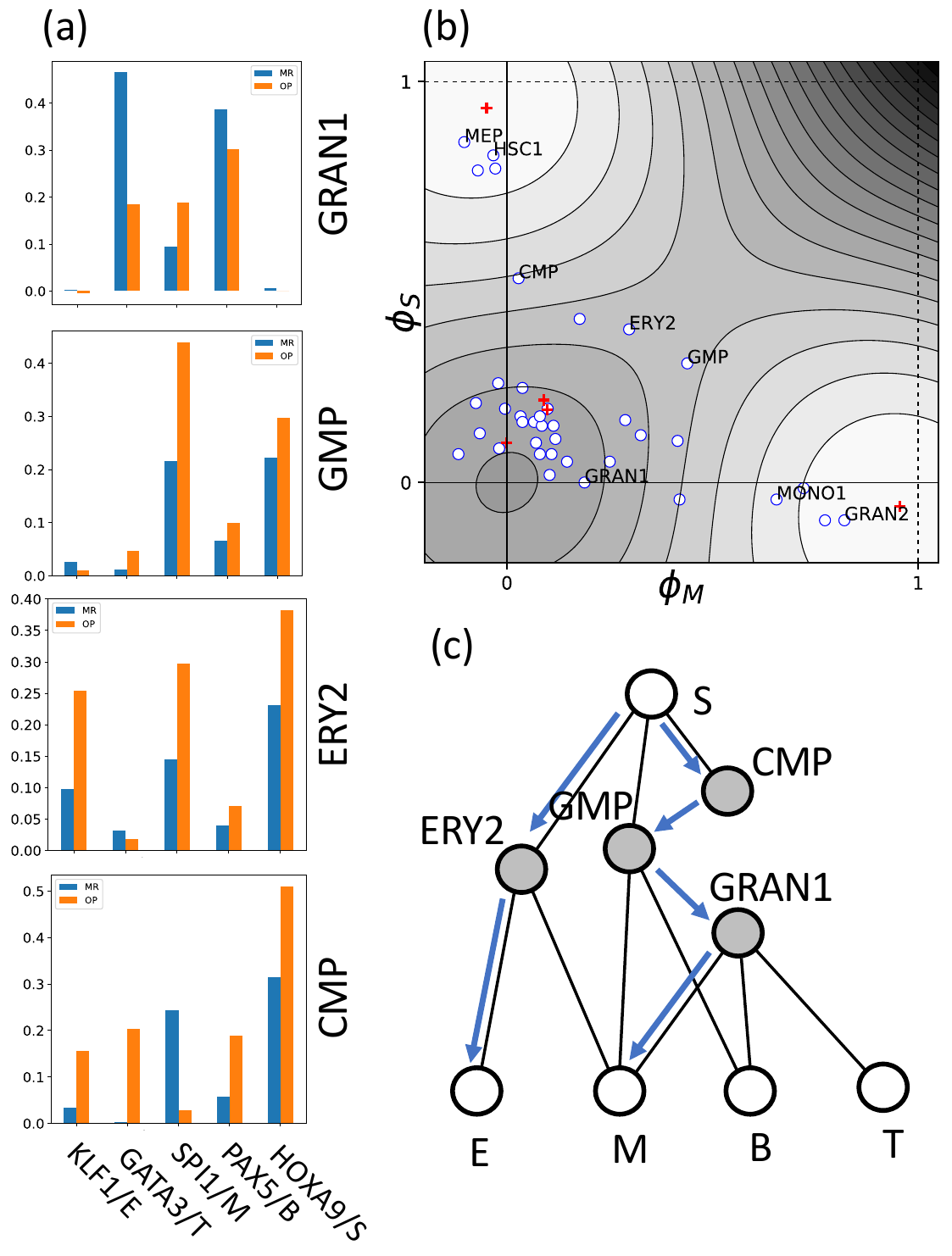} 
\caption{\label{fig3} (a) Comparison of order parameters and corresponding 
master regulators for mixed cell types GRAN1 (neutrophil progenitor), 
ERY2 (erythrocyte progenitor), CMP (common myeloid progenitor), and GMP 
(granulocyte/monocyte progenitor). (b) Cell types projected into the 
$(\phi_L, \phi_S)$-plane together with the potential $V$ in that plane. 
Projected minima of $V$ are shown as (+). (c) Graph illustrating the 
relationships of mixed cell types with the five cell categories S, T, B, M, 
and E. Arrows indicate the direction of development during hematopoiesis.}
\end{figure}

To apply the Hopfield model to this data, we need to first choose a meaningful
set of regulated genes for which patterns $\xi_i^k$ will be constructed. This is
not a trivial task because many genes are strongly co-regulated, and others do
not change their expression state across the experimental samples. The idea is
to find a gene set that is minimal in some sense but maximizes information about
which cell category a sample belongs to. For the analysis here, the selection of
genes $i$ was based on a heuristic that (a) controls the variance $\sigma^2$
across samples using a parameter $\lambda$, $\sigma^2(e_i)>\lambda$, and (b)
subsequently maximizes independence by imposing a constraint on the Pearsson
correlation coefficients $r$ controlled by a parameter $\mu$,
$\left<|r(e_i,e_j)|\right>_{j\neq i}<\mu$, where $i$ is fixed and the average
runs over all other genes. Best parameter choices $\lambda$ and $\mu$ were
determined by inspecting the sample correlation matrix based on the selected
gene set, and requiring optimal separation of clusters corresponding to the
categories S, T, B, M, and E. For illustration, this correlation matrix is shown
in Fig.~\ref{fig2}a after hierarchical clustering for $\lambda=0.25$ and
$\mu=0.30$. The size of the resulting set of regulated genes is $N=424$, where
the MR genes discussed below were also excluded for consistency.

The cell types representing the five cell categories described above are found
to be central to the clusters in Fig.~\ref{fig2}. These are HSC1,2, MEP, ERY1
for hematopoietic stem cells, TCELL2-4,6-8 for T-lymphocytes, BCELLa1-4 for
B-lymphocytes, MONO1,2, GRAN1,2 for myeloid cells, and ERY3-5 for erythroid
cells. Pattern vectors $\xi^k$ for each category $k=S, T, B, M, E$ were
constructed by averaging mapped gene expression values over these cell types,
and then applying the sign-function. It turns out that the resulting pattern
vectors are almost orthogonal, with values of
$\left<\!\left<\xi^l\xi^k\right>\!\right>$, $k\neq l$, ranging from -0.057 to
0.16, and slightly biased, with values of $\left<\!\left<\xi^k\right>\!\right>$
between -0.32 and 0.21. The model is therefore reasonably close to the symmetric
case discussed in the previous section and the condition $\phi_k\geq 0$ can be
approximately fulfilled. Minima of the resulting potential $V(\{\phi_k\})$ were
computed numerically. For simplicity, all parameters $\beta_k$, $k\in\{S, T, B,
M, E\}$, were set to the same value $\beta$. Gradually increasing $\beta$
starting from zero shows that for $\beta\gtrsim 1.5$ the point $\phi_k=0$
becomes unstable, and for $\beta\gtrsim 2$ five minima of $V$ exist, each with
one dominating OP, and small contributions of the other OPs mixed in.

For reproducibility the source code for the complete analysis is available on github \cite{supl}.

\subsection{Results}

Scaled OPs (in units of $\beta_k$),
$\phi_k^{(n)}=\frac{1}{N}\sum_is_i^{(n)}\xi_i^k$ were computed from the
expression values for each cell type $n$, where
$s_i^{(n)}=\mbox{sign}(e_i^{(n)})$. The results are compared to observed gene
expression values (as proxy for their concentration) of known MRs for the
different cell categories, HOXA9, GATA3, PAX5, SPI1, and KLF1. HOXA9 promotes
hematopoietic commitment of embryonic stem cells \cite{Ramos-Mejia2014}, GATA3
is a MR for TH2 differentiation and controls T cell maintenance and
proliferation \cite{Wang2013}, the transcription factor PAX5 is the main driver
of B cell development \cite{Medvedovic2011, Cobaleda2007}, SPI1 (also known as
PU.1) plays a crucial role in myeloid cell development \cite{Burda2010,
Friedman2007, Zakrzewska2010}, and KLF1, as one of the core erythroid
transcription factors, regulates the development of erythroid cells from
progenitors \cite{Tallack2012, Love2014}. MR expression values (not log2
transformed) were linearly mapped to the range [0, 1]. Fig.~\ref{fig2} shows
that both, OPs (Fig.~\ref{fig2}c), and MR expression (Fig.~\ref{fig2}b)
correlate well with the clusters in Fig.~\ref{fig2}a, and the cell types
defining the categories S, T, B, M, and E. 

Apart of cell types that are dominated by a single OP and MR, there are also
intermediate types corresponding to different progenitor cells particularly for
the myeloid and erythroid branches. Fig.~\ref{fig3}a shows comparisons of MR
expression and OP values for the cell types CMP (common myloid progenitor), GMP
(granulocyte/monocyte progenitor), GRAN1 (neutrophil progenitor) and ERY2
(erythrocyte progenitor). The order parameters $\phi_L$ and $\phi_S$ for these
cell types are also shown in Fig.~\ref{fig3}b, together with the potential
$V(\{\phi_k\})$ in the ($\phi_L, \phi_S$)-plane. Except for CMP (which lacks an
OP contribution corresponding to the myeloid MR SPI1), there is a good agreement
between the observed patterns of MRs and OPs. This is graphically shown in
Fig.~\ref{fig3}c, which places those cell types in the context of the categories
S, T, B, M, and E, consistent with the direction of cell development from
hematopoietic stem cells to the mature cell types. A few other cell types
(especially megakaryocytes and dendritic cells) do not fit well into the
picture, possibly because those need to be described with additional patterns
and MRs. Overall these results show that the model proposed here is consistent
with gene expression data for hematopoietic cells.\\

\section{Conclusion}

Biological systems involve a myriad of interacting components, and are too
complicated to be understood in terms of first principles. Therefore, there is
clearly a need for the development of phenomenological models abstracting from
underlying biomolecular details. In this paper, I have proposed a biologically
plausible model for cell type-specific states, in which genes act collectively
rather than in simple circuits through concurrent feedback loops. The model is
equivalent to a Hopfield model with effective Hebbian interactions, where
concentrations of master regulators are interpreted as order parameters, and an
epigenitic landscape arises as their free energy. Despite its simplifying
assumptions, it was shown that this model is consistent with experimental gene
expression data in the context of hematopoiesis. 

The model has several features that make it attractive: {\em Robustness.}
Barriers separating stable states are of the order of the system size $N$
\cite{Amit1985a}. Thus cell states are stable against fluctuations involving few
genes. {\em Parallelism.} The information transmitted through the feedback loop
involves many genes in parallel that are concurrently used by different
regulators. This could be a prototype for intracellular communication also in
other contexts since many genes are known to be shared among various cellular
functions. Parallel signalling, as long as different signals are orthogonal,
ensures that cross-talk is limited. {\em Evolvability}. The model decouples cell
type-specific patterns through separate master regulators. One may therefore
speculate, that evolution driving the ``learning'' of patterns via correlation
between forward and backward regulation can occur independently for different
cell types. Thus, multicellular organisms could evolve by adding more cell types
without perturbing existing ones. Finally, as in the case of DNA and the genetic
code for proteins, biological systems have to store information encoding their
structure and function on every level. Mapping cell states onto the Hopfield
model, which represents a prototypical information storage device, makes this
explicit.


\begin{thebibliography}{25}
\bibitem [{not()}]{note_1}%
  \BibitemOpen
  \href@noop {} {}\bibinfo {note} {Here we consider the slightly more general
  case where patterns have different weights. The parameters $\beta_k$ can also
  be interpreted as pattern-specific inverse temperatures.}\BibitemShut {Stop}%
\end{thebibliography}

%

\end{document}